\documentclass[twocolumn,aps,prc,superscriptaddress,showpacs]{revtex4}
\usepackage{multirow}
\usepackage{amssymb}
\usepackage{amsmath}
\usepackage{graphicx}
\usepackage[normalem]{ulem}
\usepackage{multirow}
\usepackage{CJK}
\usepackage[usenames]{color}

\begin{document}
\begin{CJK*} {UTF8} {gbsn}

\title{Effects of $P_{\mathrm{tot}}$ gates and velocity gates on light-particle
momentum correlation in intermediate-energy heavy-ion collisions}

\author{Ting-Ting Wang}
\affiliation{Key Laboratory of Nuclear Physics and Ion-Beam Application (MOE), Institute of Modern Physics, Fudan University, Shanghai 200433, China}
\affiliation{Shanghai Institute of Applied Physics, Chinese Academy of Sciences, Shanghai 201800, China}
\affiliation{University of Chinese Academy of Sciences, Beijing 100049, China}

\author{Yu-Gang Ma\footnote{Corresponding author: mayugang@cashq.ac.cn}}
\affiliation{Key Laboratory of Nuclear Physics and Ion-Beam Application (MOE), Institute of Modern Physics, Fudan University, Shanghai 200433, China}
\affiliation{Shanghai Institute of Applied Physics, Chinese Academy of Sciences, Shanghai 201800, China}
\affiliation{University of Chinese Academy of Sciences, Beijing 100049, China}

\author{Zheng-Qiao Zhang}
\affiliation{Shanghai Advanced Research Institute, Chinese Academy of Sciences, Shanghai 201210, China}
\affiliation{Shanghai Institute of Applied Physics, Chinese Academy of Sciences, Shanghai 201800, China}

\date{\today}

\begin{abstract}
Momentum correlation functions at small relative momenta are calculated for light particles $\left(n, p, d, t\right)$ emitted from $^{197}$Au + $^{197}$Au collisions at different impact parameters and beam energies within the framework of the isospin-dependent quantum molecular dynamics model complemented by the $Lednick\acute{y}$ and $Lyuboshitz$ analytical method.
We first make sure our model is able to reproduce the FOPI data of proton-proton momentum correlation in a wide energy range from 0.4$A$ GeV to 1.5$A$ GeV. Then we explore more physics insights through the emission times and momentum correlations among different light particles. The specific   emphasize is the effects of total pair momentum among different light particles, impact parameters and in-medium nucleon-nucleon cross section.   Both two-deuteron and two-triton correlation functions are anti-correlation due to the final state interaction, and they are affected by in-medium nucleon-nucleon cross section for the higher total momentum of the particle pairs, but not for the lower ones. In addition, impact parameter and in-medium nucleon-nucleon cross section dependences of the emission source radii are extracted by fitting the momentum correlation functions. The results indicate that momentum correlation functions gating with total pair momentum is stronger for the smaller in-medium nucleon-nucleon cross section factor $\left(\eta\right)$ or impact parameter $\left(b\right)$. Non-identical particle correlations ($np, pd, pt, $ and $dt$) are also investigated by the velocity-gated correlation functions which can give information of the particles' emission sequence, and the result indicates that heavier ones $\left(deuteron/triton\right)$ are, one the average, emitted earlier than protons, in the small relative  momentum region.
\end{abstract}
\pacs{25.70.Mn, 24.10.-i, 25.70.Pq, 27.80.+w}

\maketitle

\section{Introduction}
It is well known that two-particle momentum correlation function at small relative momenta is sensitive to the space-time structure of  particles at the freeze-out, and therefore the characteristics of the particle emission source
\cite{Kopylov1973,Kopylov1974,Koonin1977,Pratt1984,Pratt1986,Pratt1987,Boal1990,Kotte1999,Orr,Gong1990,Pochodzalla1986,Ma2015PLB,Fang2016PRC}. At relativistic energy, two-particle correlation function  provides a very useful tool for measuring the freeze-out properties of  the partonic or  hadronic systems \cite{Lisa,Star1,Phenix1,LinZW,Yang2018,Liu2018} as well as interaction parameters between particle pairs \cite{Star-nature,p-omega,p-lambda,lambda-lambda,Chen}.
In intermediate-energy region, the two-proton correlation function was mostly taken as a probe for the space-time properties such as the source size and emission time in  nuclear reactions \cite{Kotte2005,Ghetti2003,Gourio2000}. Many investigations of the two-proton correlation function have been done by a lot of experiments
and explored by different models, including various effects of the impact parameter \cite{Gong1991,Ma2006}, the total momentum of nucleon pairs \cite{Ma2006}, the isospin of the emission source \cite{Ghetti2004}, the nuclear symmetry energy \cite{Chen2003}, the nuclear equation of state (EOS) \cite{Ma2006}, and the in-medium nucleon-nucleon cross section (NNCS) \cite{Ma2006,wtt2018} etc. Particularly, the dependence of the two-proton correlation function on the in-medium NNCS has been studied in more details via the Pratt's CRAB code \cite{Ma2006} or the Lednick$\acute{y}$ and Lyuboshitz  code \cite{wtt2018,Zhang2007} in the framework of an isospin-dependent quantum molecular dynamics (IQMD) model. Since the magnitude of the total pair momentum is related to the nucleon emission time, the effect of total nucleon pair momentum on the strength of the correlation function was also discussed in heavy-ion collisions \cite{Chen2003npa,Chen2003,Ma2006}.

The correlation functions between two light charged particles other than two protons carry more information about the light particle production mechanism and reaction dynamics in heavy-ion collisions at intermediate energies \cite{Koonin1977,Pratt1987,Boal1990,Gong1991,Cao2012,Chitwood1985,Pochodzalla,Kryger1990,Zhu1991,Gong1993,Erazmus1994,Hamilton1996,Verde2006}. In previous work \cite{Pochodzalla,Erazmus1994,Cebra1989,Boal1986}, it is demonstrated that source sizes extracted from different particle species correlation functions are different. It may be attributed to the dynamical expansion of the reaction zone and different time scales \cite{Gelderloos1994}. The simultaneous investigation of correlation functions involving composite light particles may offer a unique tool to investigate dynamical expansion of the reaction zone \cite{Boal1990}.

On the other hand, momentum correlations between two non-identical particles contain information on the emission time differences of the two particles. Therefore, by comparing the correlation functions between two non-identical particles  with different velocity gates, one could infer the emission sequence between these two non-identical particles \cite{Gelderloos1995,Lednicky1996,Voloshin1997},
such as $p$, $d$, $t$, $^{3}\textrm{He}$ and so on  \cite{Kotte1999,Voloshin1997,lednicky1982}.

In this paper, we will discuss the correlation functions of the light particles at different centralities and the in-medium nucleon-nucleon cross section. In addition, we will investigate correlation functions of light particles under different total momentum of particle  pairs. Furthermore, we also like to check  whether the strength of the correlation functions for light particle pairs with higher/lower total pair momenta is sensitive to the in-medium nucleon-nucleon cross section. On the other hand, for the two non-identical light particle pairs, we can get information about the  order of emission time from their correlation functions gated with velocity selection. We have applied this method to the $n-p/p-d/p-t/d-t$ correlation functions for particles emitted in lower relative momentum region.

To study the above questions quantitatively, a theoretical approach proposed by Lednick$\acute{y}$ and Lyuboshitz  \cite{lednicky2006}  is applied for momentum correlation function construction based on the phase space data by  an isospin-dependent quantum molecular dynamics  (IQMD) model. To this end, we use $^{197}$Au + $^{197}$Au system to investigate momentum correlation functions at different  beam energies and impact parameters.

The rest of paper is organized as follows. In Section 2, we briefly describe the Hanbury-Brown Twiss (HBT) technique using the $Lednick\acute{y}$ and $Lyuboshitz$ analytical formalism and an isospin-dependent quantum molecular dynamics model. In Section 3, we show
the results of the IQMD plus the Lednick$\acute{y}$ and Lyuboshitz method for the study of proton-proton correlation function, where  the results are compared with the FOPI experimental data. We then  systematically discuss light particle momentum correlation function and the influences of gates on the total momentum of the light particle pairs. The detailed analysis of light particle momentum correlation functions and extracted source size results are presented under  different in-medium nucleon-nucleon cross sections and  impact parameters for Au + Au collisions. Furthermore,  correlation functions of non-identical light particles are analyzed to deduce the emission time order of the two different particles in lower relative momentum region. In the end,  Section 4 gives a summary of the article.

\section{FORMALISM AND MODELS}

\subsection{LEDNICK$\acute{Y}$ and LYUBOSHITZ ANALYTICAL FORMALISM}

Firstly, we  present a brief review of a theoretical approach given by Lednick$\acute{y}$ and Lyuboshitz \cite{lednicky1982,lednicky2006,Lednicky2009,Lednicky2008} for the HBT technique and the understanding of physics in the present work.
In such a framework, the main formula is based on the principle that the particle correlations when they are emitted at small relative momenta are determined by the space-time characteristics of
the production processes owing to the effects of quantum statistics (QS) and final-state interactions (FSI) \cite{Koonin1977}. Then, the correlation function can be expressed through a square of the symmetrizied Bethe-Salpeter amplitude averaging over the four coordinates of the emission particles and the total spin of the two-particle system, which represents the continuous spectrum of the two-particle state.
In this model, the FSI of the particle pairs is assumed independent in the production process. According to the conditions in Refs. \cite{Lednicky1996}, the correlation function of two particles can be written as the expression:
\begin{equation}
\textbf{C}\left(\textbf{k}^*\right) = \frac{\int
\textbf{S}\left(\textbf{r}^*,\textbf{k}^*\right)
\left|\Psi_{\textbf{k}^*}\left(\textbf{r}^*\right)\right|^{2}d^{4}\textbf{r}^*}
{\int
\textbf{S}\left(\textbf{r}^*,\textbf{k}^*\right)d^{4}\textbf{r}^*},
\end{equation}
where $\textbf{r}^* = \textbf{x}_{1}-\textbf{x}_{2}$ is the relative distance of the two particles at their kinetic freeze-out, $\textbf{k}^*$ is half of the relative momentum between two particles, $\textbf{S}\left(\textbf{r}^*,\textbf{k}^*\right)$ is the probability to emit a particle pair with given $\textbf{r}^*$ and $\textbf{k}^*$, $i.e.$, the source emission function, and $\Psi_{\textbf{k}^*}\left(\textbf{r}^*\right)$ is Bethe-Salpeter amplitude which can be approximated by the outer solution of the scattering problem \cite{Star-nature}.
In above limit, the asymptotic solution of the wave function of the two charged particles approximately takes the expression:
\begin{multline}
\Psi_{\textbf{k}^*}\left(\textbf{r}^*\right) = e^{i\delta_{c}}\sqrt{A_{c}\left(\lambda \right)} \times\\
\left[e^{-i\textbf{k}^*\textbf{r}^*}F\left(-i\lambda,1,i\xi\right)+f_c\left(k^*\right)\frac{\tilde{G}\left(\rho,\lambda \right)}{r^*}\right].
\end{multline}
In the above equation, $\delta_{c} =
$arg$\Gamma\left(1+i\lambda\right)$ is the Coulomb s-wave phase shift with $\lambda = \left(k^*a_c\right)^{-1}$ where $a_{c}$ is the two-particle Bohr radius, $A_c\left(\lambda \right) = 2\pi\lambda \left[\exp\left(2\pi\lambda \right)-1\right]^{-1}$ is the Coulomb penetration factor, and its positive (negative) value corresponds to the repulsion (attraction). $\tilde{G}\left(\rho,\lambda \right) = \sqrt{A_{c}\left(\lambda \right)}\left[G_0\left(\rho,\lambda \right)+iF_0\left(\rho,\lambda \right)\right]$ is a combination of regular $\left(F_0\right)$ and singular $\left(G_0\right)$ s-wave Coulomb functions \cite{Lednicky2009,Lednicky2008}. $F\left(-i\lambda,1,i\xi\right) = 1+\left(-i\lambda\right)\left(i\xi\right)/1!^{2}+\left(-i\lambda\right)\left(-i\lambda+1\right)\left(i\xi\right)^{2}/2!^{2}+\cdots$ is the confluent hypergeometric function with $\xi = \textbf{k}^*\textbf{r}^*+\rho$, $\rho=k^*r^*$.

\begin{equation}
f_c\left(k^*\right) = \left[ K_{c}\left(k^*\right)-\frac{2}{a_c}h\left(\lambda \right)-ik^*A_{c}\left(\lambda \right)\right]^{-1}
\end{equation}
is the s-wave scattering amplitude renormalizied by the long-range Coulomb interaction, with $h\left(\lambda \right) = \lambda^{2}\sum_{n=1}^{\infty}\left[n\left(n^2+\lambda^2\right)\right]^{-1}-C-\ln\left[\lambda \right]$ where $C$ = 0.5772 is the Euler constant.
$K_{c}\left(k^*\right) = \frac{1}{f_0} + \frac{1}{2}d_0k^{*^2} + Pk^{*^4} + \cdots$ is the effective range function, where $d_{0}$ is the effective radius of the strong interaction, $f_{0}$ is the scattering length and $P$ is the shape parameter. The parameters of the effective range function are important parameters characterizing the essential properties of the FSI, and can be extracted from the correlation function measured experimentally \cite{Erazmus1994,Arvieux1974,Star-nature}.
Table I shows the parameters of the effective range function for different particle pairs in the present work.

\begin{table}[!htbp]
\caption{Experimental determination of the effective range function parameters for n-n, p-p, t-t, p-d, p-t, d-t and n-p systems \cite{Erazmus1994,Star-nature,Arvieux1974,Landau1974}.}
\begin{tabular}{ccccc}
\hline
\hline
\multicolumn{1}{r}{System} & Spin & $f_{0}$ $\left(fm\right)$ & $d_{0}$ $\left(fm\right)$ & $P$ $\left(fm^{3}\right)$ \\ \hline
n-n                          & 0    & 17     & 2.7     & 0.0     \\
p-p                          & 0    & 7.8     & 2.77     & 0.0     \\
t-t                          & 0    & $1*10^{-6}$  & 0.0      & 0.0     \\
\multirow{2}{*}{p-d}         & 1/2  & -2.73   & 2.27     & 0.08    \\
                            & 3/2  & -11.88  & 2.63     & -0.54   \\
p-t                          & 0    & $1*10^{-6}$  & 0.0      & 0.0     \\
d-t                          & 0    & $1*10^{-6}$  & 0.0      & 0.0     \\
n-p                          & 0    & 23.7  & 2.7      & 0.0     \\ \hline
\hline
\end{tabular}
\end{table}

In the above table, for $n-n$ and $n-p$ correlation functions which include uncharged particle, the Coulomb penetration factor ($A_c\left(\lambda \right)$) is not considered and only the short-range particle interaction works. For charged particles correlation functions, only effect of the Coulomb interaction is expected to dominate the correlation functions of $t-t$, $p-t$, and $d-t$ system. However, except the Coulomb interaction, the short-range particle interaction dominated by the s-wave interaction is considered for $p-p$, $d-d$ and $p-d$ particle pairs at the small relative momenta. The correlation function of $p-p$ particle pairs is dominated by only the singlet $\left(S = 0\right)$ s-wave FSI contribute while both spins 1/2 $\left(doublet\right)$ and 3/2 $\left(quartet\right)$ contribute in the case of $p-d$ system.
However, for deuteron-deuteron correlation function, a parametrization of the s-wave phase shifts $\delta$ has been used from the solution of $K_{c}\left(k^*\right) = \cot{\delta}$ for each total pair spin $S = 0, 1, 2$. Note that the effective range function for the total spin $S = 1$ is irrelevant, since it does not contribute due to the QS symmetrization.

\subsection{THE IQMD MODEL}

In a specific application of the Lednick$\acute{y}$ and Lyuboshitz theoretical simulation, the true single-particle phase-space distribution at the freeze-out stage is required. In this paper, the isospin-dependent Quantum Molecular Dynamics  transport model is used as the event generator, which has been applied successfully to the HBT studies   in intermediate-energy heavy-ion collisions  (HICs) \cite{Ma2006,Ma2007,Wei2004,Wei2004jpg,Cao2012,wtt2015,wtt2018}.
In the following discussion, we introduce the model briefly. The Quantum Molecular Dynamics transport model is a n-body transport theory, it describes heavy-ion reaction dynamics from intermediate to relativistic energies \cite{Aichelin1987,Aichelin1991,Peilert1989,Li2018,Feng2018}.  Since the QMD transport model contains correlation effects for all orders, one can investigate various information on both the collision dynamics and the fragmentation process \cite{MaCW,Yan,wtt2015,wtt2018}. The main  parts of QMD transport model include the following issues: the initialization of the projectile and the target, nucleon propagation under the effective potential, the collisions between the nucleons in the nuclear medium, the Pauli blocking effect, and the numerical tests.

The isospin-dependent Quantum Molecular Dynamics transport model is based on the QMD transport model with the isospin factors. As we know, the main components of the dynamics in HICs at intermediate energies include the mean field, two-body collisions, and Pauli blocking. Therefore, it is important for these three components to include isospin degree of freedom in the IQMD transport model. What is more, due to a large difference between neutron and proton density distributions for nuclei far from the $\beta$-stability line, the samples of neutrons and protons in phase space should be treated separately in the projectile and target nuclei initialization.

In the IQMD model, the interaction potential is represented by the form as follows:
\begin{equation}
U = U_{Sky} + U_{Coul} + U_{Yuk} + U_{Sym} + U_{MDI} + U_{Pauli},
\end{equation}
where $U_{Sky}$, $U_{Coul}$, $U_{Yuk}$, $U_{Sym}$, $U_{MDI}$,
and $U_{Pauli}$
are the density-dependent Skyrme potential, the Coulomb potential, the surface Yukawa potential, the isospin asymmetry potential, the momentum-dependent interaction and the Pauli potential, respectively.

In particular,  the density-dependent Skyrme potential $U_{Sky}$
reads when the momentum dependent potential is included
\begin{equation}
U_{Sky}=\alpha (\frac \rho {\rho _0})+\beta (\frac \rho {\rho
_0})^\gamma + t_4 ln^2[\varepsilon(\frac \rho {\rho
_0})^{2/3}+1]\frac \rho {\rho _0},
\end{equation}
where $\rho$ and $\rho _{_0}$ are total nucleon density and its
normal value, respectively. The parameters $\alpha$, $\beta$,
$\gamma$, $t_4$ and $\varepsilon$ are related to the nuclear
equation of state \cite{BALi,Zhang-Li17,Cai17,Zhang17} and listed in Table II,
 where $K$ = 200 or 380 MeV means  the soft- or
the stiff- momentum dependent potential, respectively.

\begin{table}[!htbp]
\caption
{The parameters of the interaction potentials.}
\begin{tabular}{|c|c|c|c|c|c|c|c|}
\hline
$\alpha$ & $\beta$&$\gamma$&$t_{4}$& $\varepsilon$ & K\\
\hline (MeV)&(MeV)&&(MeV)&(MeV)&(MeV)\\\hline
-390.1&320.3&1.14&1.57&21.54&200\\\hline
-129.2 &59.4&2.09 &1.57&21.54&380\\\hline
\end{tabular}\\
\end{table}

A general review of the above potentials can be found in Ref. \cite{Aichelin1987}.
In the present work, the in-medium nucleon-nucleon cross section with isospin-dependence is represented by the formula:

\begin{equation}
\sigma_{NN}^{med} = \left(1-\eta\frac{\rho}{\rho_{0}}\right)\sigma_{NN}^{free},
\end{equation}
where $\rho_{0}$ is the normal nuclear matter density, $\rho$ is the local density, $\eta$ is the in-medium  NNCS factor and $\sigma_{NN}^{free}$ is the available experimental NNCS \cite{Chen1968}. The above reduction factor of the in-medium NNCS was introduced by the studies of collective flow in HICs at intermediate energies \cite{Greiner1993,Klakow1982,Ma1995}. In particularly, the factor $\eta \approx 0.2$ has been found better to reproduce the flow data.

In this model, the particles are identified using a modified isospin-independent coalescence description, $i.e.$, Minimum Spanning Tree approach. In the Minimum Spanning Tree approach, nucleons are assumed to share the same cluster if their centers are closer than a distance of 3.5 $fm$ and their relative momentum smaller than 0.3 GeV/$c$.
In the present calculations, protons and neutrons are considered to be emitted when its surrounding density falls below a value of 0.02$/fm^{3}$ and unbound protons and neutrons for which no other nucleon exists within a coalescence distance of 3.5 $fm$ and relative momentum smaller than 0.3 GeV/$c$ before the freeze-out time.
If the nucleon is not bounded by any clusters, it is treated by an emitted (free) nucleon. In our calculations, the reactions of $^{197}$Au + $^{197}$Au are performed. We use the soft EOS with momentum dependent interaction for different impact parameters at different beam energies.  For each run and particle species, the momentum correlation function is constructed when the system is basically at the corresponding freeze-out time and then processed within the Lednick$\acute{y}$ and Lyuboshitz model.
\begin{figure*}[!htbp]
 \includegraphics[width=\linewidth]{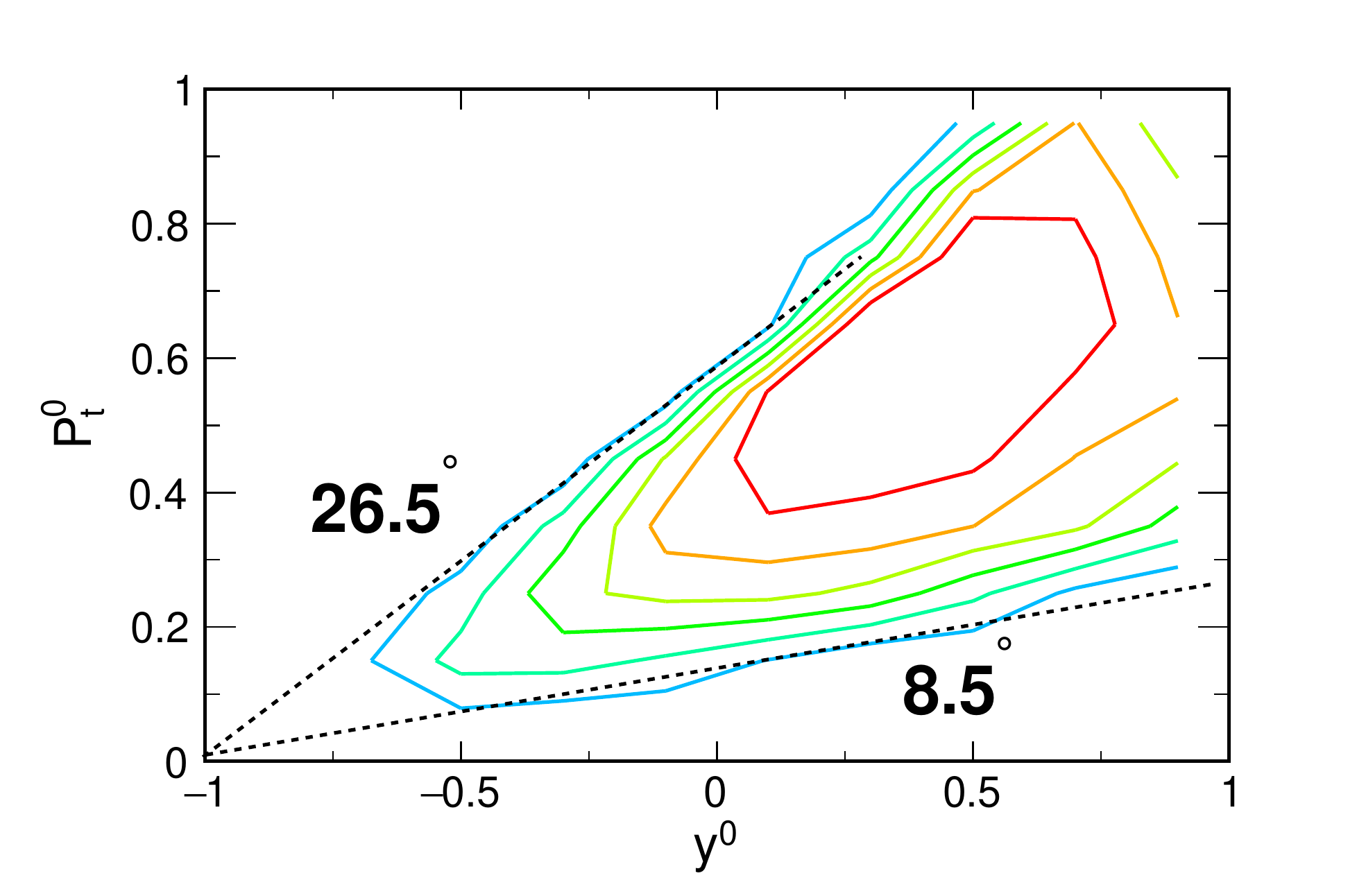}
 \centering
 \caption{(Color online)
  Two-dimensional distribution of yields of proton in the $P_{t}^{0}-y^{0}$ plane for central $^{197}$Au + $^{197}$Au reactions. Target and projectile rapidities are given by $y^{0}$ = -1 and +1, respectively. The polar angle limits at 8.5 and 26.5 degrees.
 }
 \label{fig1_pty}
\end{figure*}

\begin{figure*}[!htbp]
 \includegraphics[width=\linewidth]{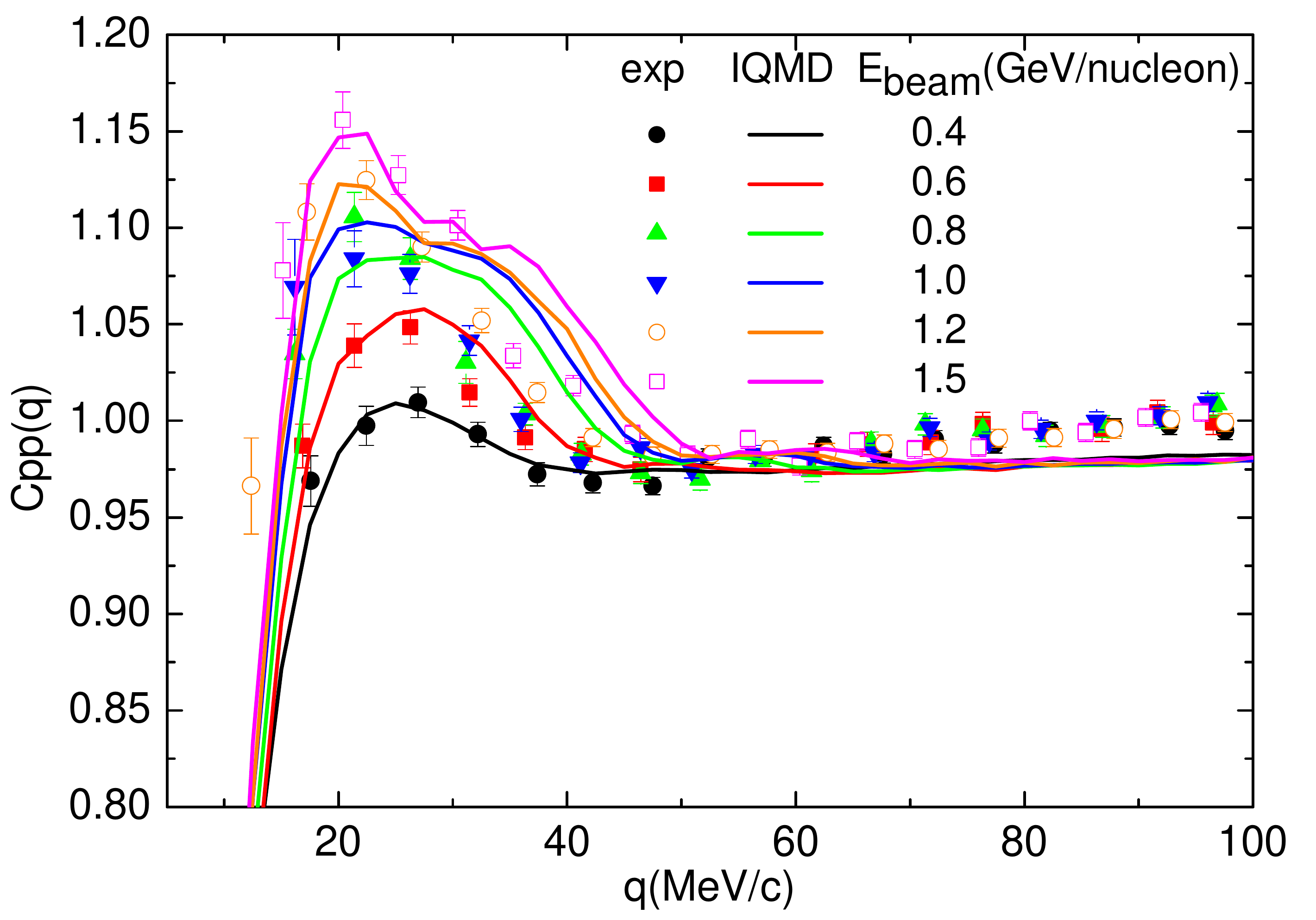}
 \centering
 \caption{(Color online)
  Proton-proton correlation functions for the central Au + Au collisions at beam energies from 0.4 to 1.5$A$ GeV. Experimental data (symbols) are compared to our predictions by the IQMD + FSI model calculation (lines).
 }
 \label{fig2_fopi}
\end{figure*}

\section{ ANALYSIS AND DISCUSSION}
\subsection{Comparison of the model predictions with experimental results}

The collision centrality is an important variable for controlling reaction dynamics. Experimentally it could be estimated by the total multiplicity distribution of charged particles
 \cite{Kotte1999}. In previous  FOPI experiments,  total multiplicity distribution was measured in the outer Plastic Wall  \cite{Kotte1999}. For a specific  selection of central collision,  the corresponding integrated cross-sections for the collision system of  Au + Au about 10\% of the total cross-section has been selected \cite{Kotte2005}. To make a quantitative comparison with  experimental data at 10\% centrality  \cite{Kotte2005}, one would use the impact parameter of about 3 $fm$  in the IQMD model for Au + Au collision and the proton is selected in the polar-angle $\left(8.5^{\circ} \leq\theta_{lab}\leq 26.5^{\circ}\right)$ triggered in the middle rapidity as  Ref.~\cite{Kotte2005} did.

 Fig.~\ref{fig1_pty} shows the phase space coverage corresponding to the experimental distributions in the c.m. system in central collisions. Here, $P_{t}^{0} = \left(p_t/A_{clus}\right)/\left(p_{proj}/A_{proj}\right)_{cm}$ and $y^{0} = \left(y/y_{proj}\right)_{cm}$ are the normalized transverse momentum and rapidity, respectively.
Within the above cut of  $P_{t}^{0}$ and $y^{0}$, we confront the experimental beam energy dependence of two-proton correlations with the predictions of the IQMD + Lednick$\acute{y}$ and Lyuboshitz hybrid model. Fig.~\ref{fig2_fopi} shows our calculated proton-proton correlation functions for central Au + Au collisions in comparison to the experimental results. In figure, $q$ at X-axis represents half of relative momentum between particle pair, i.e. $k^*$ in Eq. (1). In all following figures, $q$ is the same quantity.  With the above conditions in the transport approach the correlation functions  nicely agree with the data. We would like to point out that the fits of our correlation functions predicted by the IQMD to those from the experimental data is much better than previous correlation functions predicted by the BUU calculations \cite{Kotte2005}. With increasing beam energy the peak of the proton-proton correlation function increases, and hence the apparent source radius decreases. The trend is similar to the one that can be found in Refs. \cite{Ma2006,wtt2018}.

\begin{figure}[!htbp]
 \includegraphics[width=\linewidth]{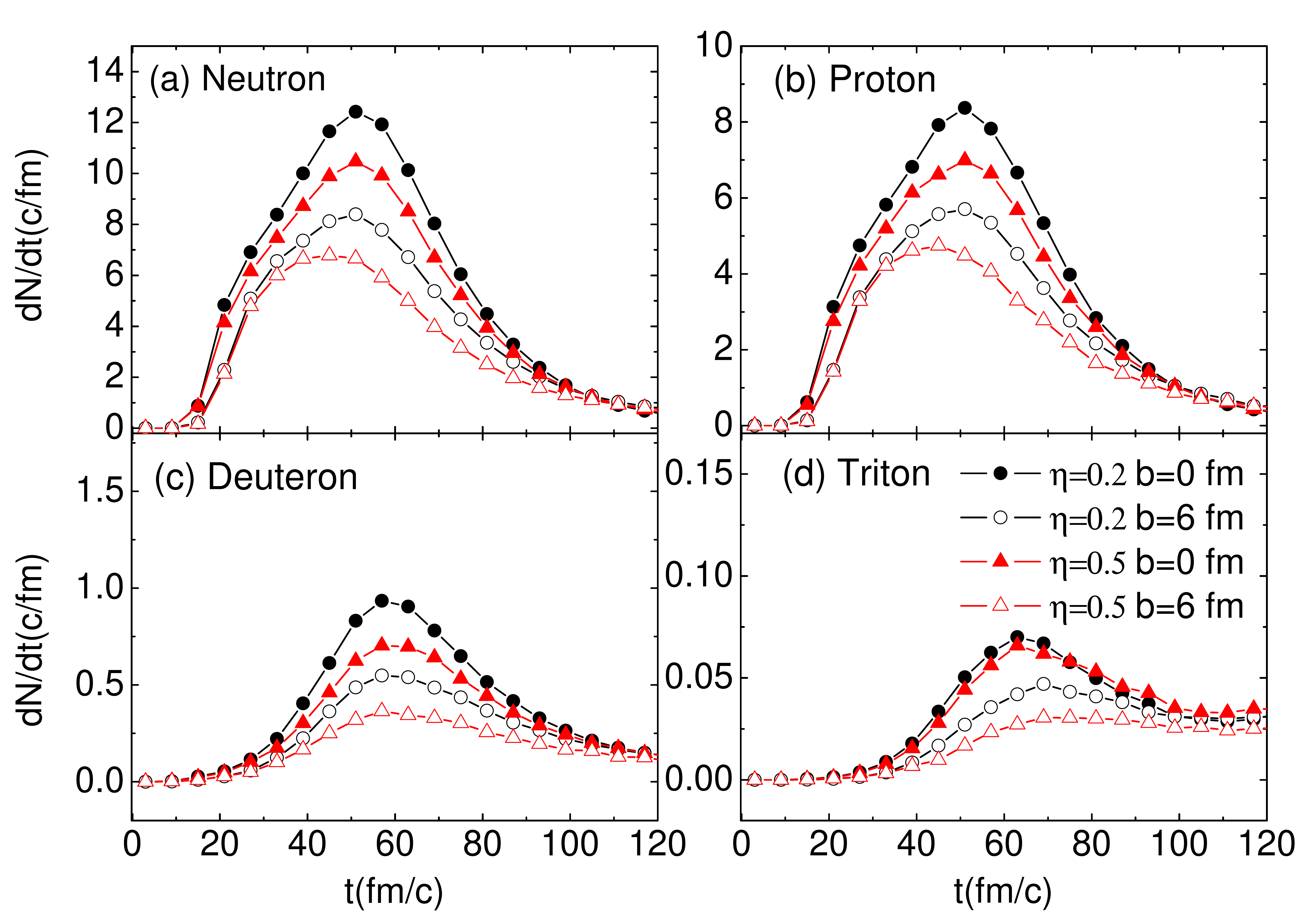}
 \centering
 \caption{(Color online)
Emission time distributions for neutrons (a), protons (b), deuterons (c), and tritons (d) for Au + Au collisions at 0.4$A$ GeV with different in-medium nucleon-nucleon cross section factor $\left(\eta = 0.2, 0.5\right)$ and impact parameters $\left(b = 0.0, 6.0 fm \right)$.
 }
 \label{fig3_emission-time}
\end{figure}

\begin{figure}[!htbp]
 \includegraphics[width=\linewidth]{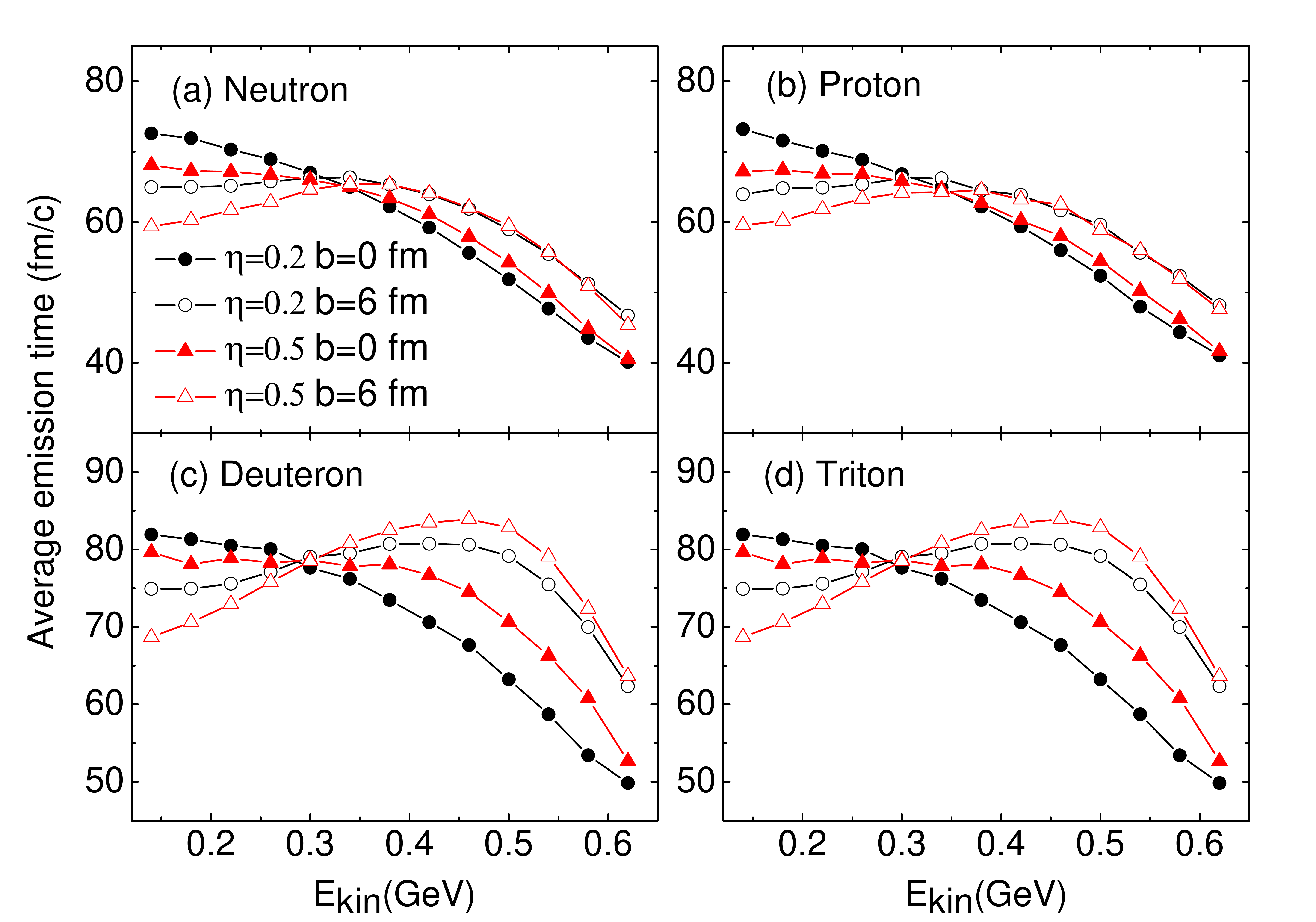}
 \centering
 \caption{(Color online)
   Average emission times of neutrons (a), protons (b), deuterons (c), and tritons (d) as a function of their c.m. kinetic energy for the Au + Au collisions at 0.4$A$ GeV.
 }
 \label{fig4_emission-ekin}
\end{figure}

\subsection{Emission times of neutrons, protons, deuterons, and tritons}

On the  basis of the good fits of proton-proton momentum correlation between the data and our calculations, we will proceed in the following sections for more detailed calculations and discussion on momentum correlation functions among neutrons, protons, deuterons and tritons, especially for investigating the effects of pair momentum cuts and the in-medium nucleon-nucleon cross section as well as the emission time sequence among neutrons, protons, deuterons and tritons.

Here we are starting from the discussion on emission time distribution of different light particles since they are relevant for understanding both the collision dynamics and the mechanism of particle production. In heavy-ion collisions at intermediate energies, nucleon emissions are mainly governed by the pressure of  excited nuclear matter during the initial stage of collisions \cite{Chen2003npa}. We performed calculations for different choices of a density dependent in-medium nucleon-nucleon cross section with the $\eta$ factors of 0.2 and 0.5, and impact parameter at b = 0.0 and  6.0 fm. In previous studies,  the choice of $\eta = 0.2$ provides the best agreement with the balance energy in collective flow data. To see the $\eta$ and impact parameter effects on light particle emissions, we show in Fig.~\ref{fig3_emission-time} $\left(a-d\right)$ the emission time distributions for neutrons, protons, deuterons and tritons, respectively, for Au + Au collisions at 0.4$A$ GeV. We can see that the emission time distribution of neutron is similar to that of protons. However, the emission time distributions of light particles are different from that of protons and neutrons. While the proton and neutron emission time peaks earlier at about 50 $fm/c$, the emission time of light particles peaks later at about 60 $fm/c$. As to the $\eta$ and impact parameter effects on particle emission, we find that the particle emission rates are larger in the cases of  smaller $\eta$ or $b$ because the larger in-medium nucleon-nucleon cross section or central collisions gives a larger initial pressure which pushes more particles emission. On the emission time, only slight differences are  there.

Particles emitted in earlier stage of heavy-ion collisions usually have higher energy than those emitted during later stage of the reaction. It is thus of interest to study the relationship between the average emission times of particles and their kinetic energy.
Shown in Fig.~\ref{fig4_emission-ekin} are the average emission times of neutrons, protons, deuterons, and tritons as a function of their c.m. kinetic energy under  the same condition as Fig.~\ref{fig3_emission-time}.  We  see that the particles with higher kinetic energies are emitted earlier than those with lower kinetic energies at central collision (i.e. b= 0 fm). However, at b = 6 fm, the average  emission times are not a monotonous function of the kinetic energy, especially for deuterons and tritons.  The above difference indicates that different emission mechanism at central collisions or semi-peripheral collisions (b = 6 fm). In central collisions, most light particles emissions are mainly driven by a high-pressure dynamical source but at semi-peripheral collisions, light particle emissions are competed by an overlapping dynamical source with thermal source.
In relative higher kinetic energy region, eg. above $\sim$ 0.32 GeV for neutrons and protons, and 0.30 GeV for deuterons and tritons, the average emission times become later when the $\eta$ becomes larger, i.e. the small in-medium nucleon-nucleon cross section.
However, in relative lower kinetic energy region, the effect of in-medium nucleon-nucleon cross section on the average emission time is just reverse.

\begin{figure}[!htbp]
 \includegraphics[width=\linewidth]{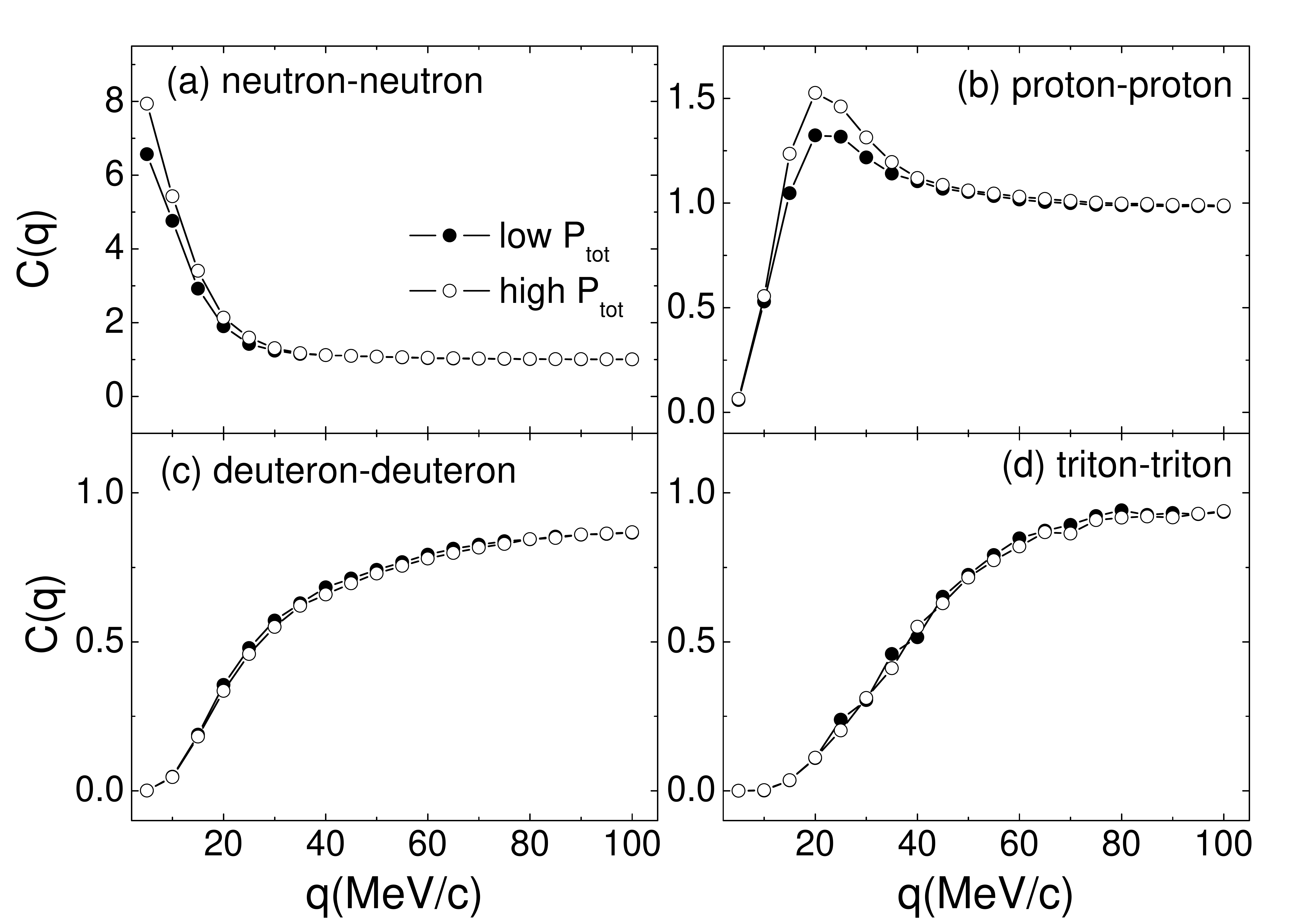}
 \centering
 \caption{
Momentum  correlation functions of particle-pairs for Au + Au central collisions at 0.4$A$ GeV with the cuts of different total particle pair momentum. Open and filled circles correspond to high $P_{tot}$ and low $P_{tot}$ cuts, respectively. (a) neutron pairs gated on  $P_{tot}$:   low: 0- 0.4 GeV/c, high: 0.8-1.2 GeV/c; (b) proton pairs gated on  $P_{tot}$:   low: 0- 0.4 GeV/c, high: 0.8-1.2 GeV/c;  (c)  deuteron pairs gated on  $P_{tot}$:  low: 0-  0.8 GeV/c, high: 1.6 - 2.4 GeV/c;  (d)  triton pairs gated on  $P_{tot}$: low: 0 - 1 GeV/c, high: 2 - 3 GeV/c, respectively.
 }
 \label{fig5_ptot}
\end{figure}

\begin{figure*}[!htbp]
 \includegraphics[width=\linewidth]{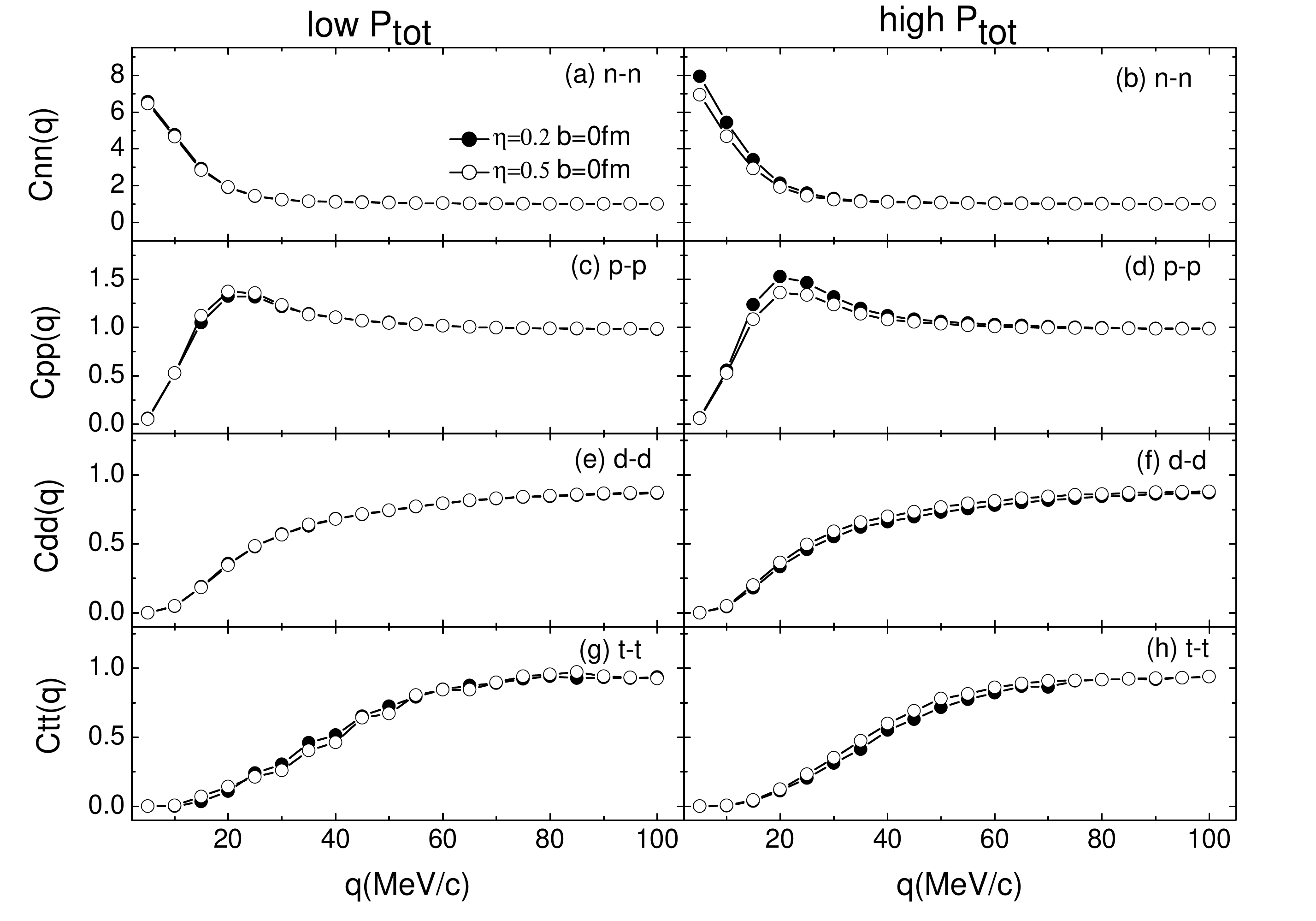}
 \centering
 \caption{
 Momentum  correlation functions of particle-pairs as a function of different $\eta$-factor in central collisions at incident energy $E$ = 0.4$A$ GeV. From upper  to bottom panel, it corresponds to  the correlation functions of neutron-neutron, proton-proton, deuteron-deuteron, and triton-triton gated on low $P_{tot}$ (left panels) and high $P_{tot}$ (right panels), respectively.
 }
 \label{fig6_ptot-eta}
\end{figure*}

\begin{figure}[!htbp]
 \includegraphics[width=\linewidth]{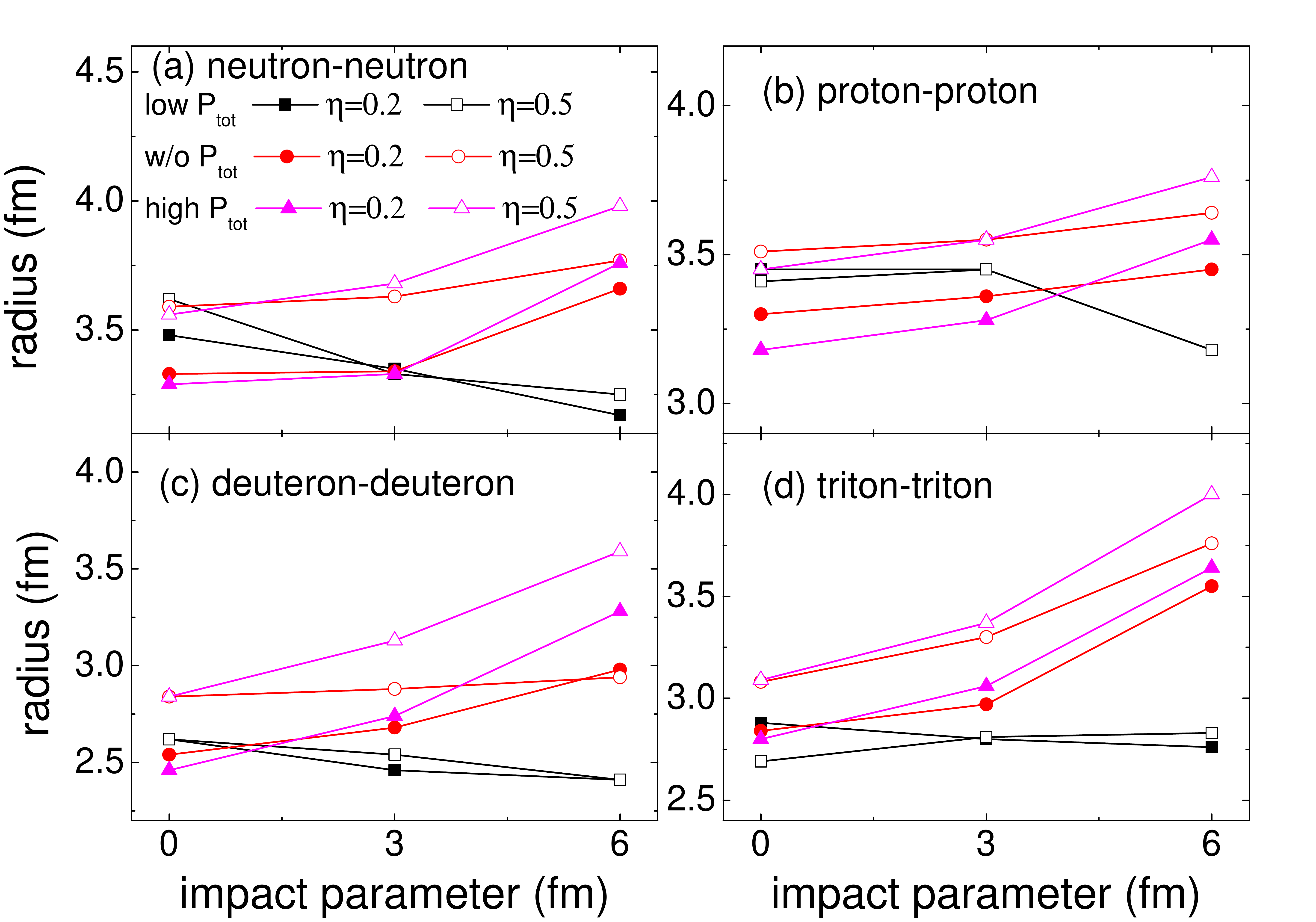}
 \centering
 \caption{(Color online)
Gaussian source radius as a function of  impact parameter at different $\eta$-factors at fixed incident energy 0.4$A$ GeV. From (a) to (d), it corresponds to  the Gaussian radius of neutrons, protons, deuteron, and triton pairs, respectively.
 }
 \label{fig7_radius}
\end{figure}

\begin{figure}[!htbp]
 \includegraphics[width=\linewidth]{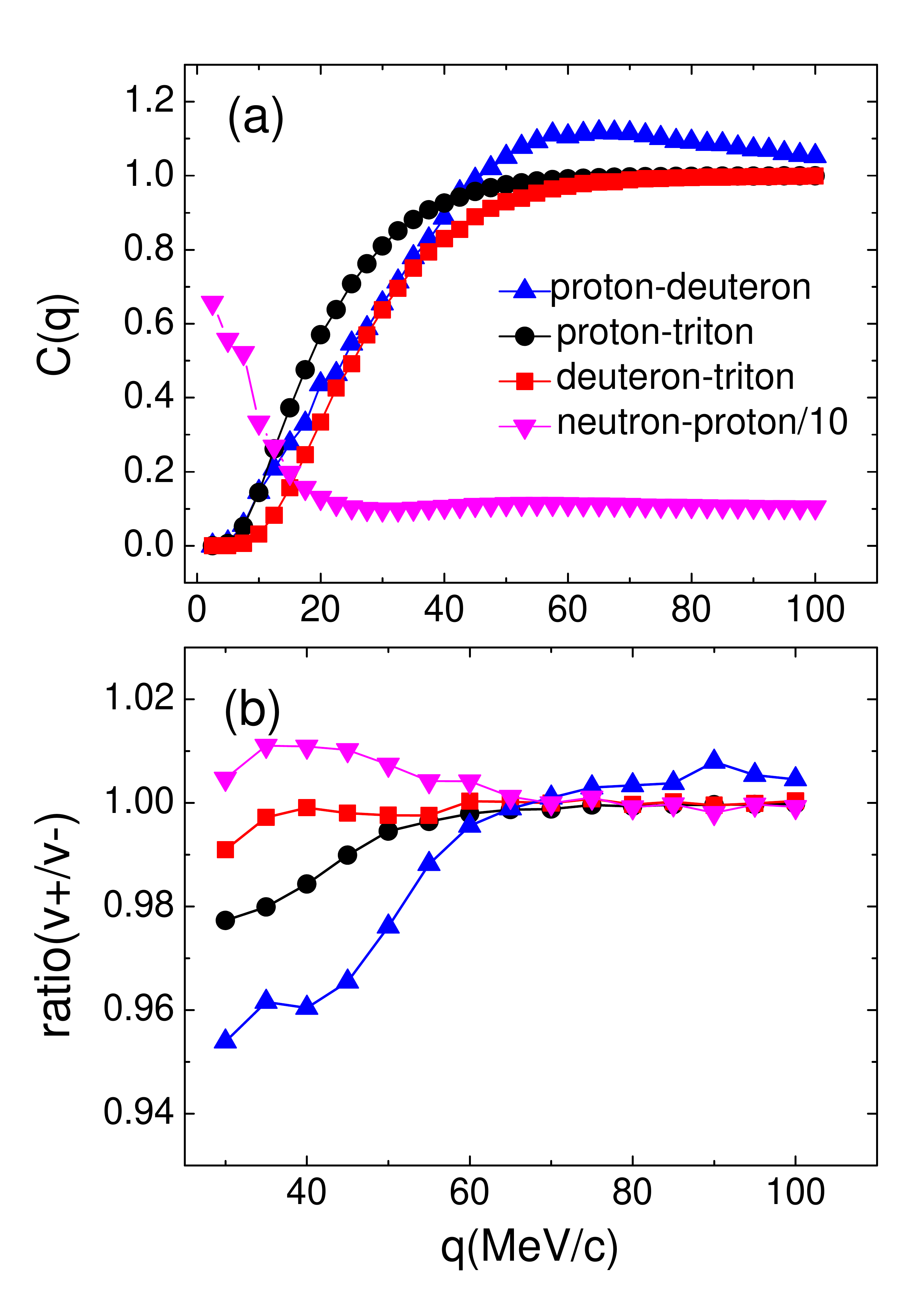}
 \centering
 \caption{(Color online)
Momentum  correlation functions of non-identical particle-pairs for central Au + Au  collisions at 0.4$A$ GeV. The upper panel corresponds to proton-deuteron pairs,  proton-triton pairs, deuteron-triton pairs and neutron-proton, respectively. The neutron-proton correlation function is scaled down 10 times. The bottom panel shows the ratio of both functions grouped into two velocity bins (v+ and v-) which v+ contains the particle faster than the other one and v- is the reverse situation. See texts for details.
 }
 \label{fig8_vpdt-ratio}
\end{figure}

\subsection{Correlation functions of neutrons, protons, deuterons, and tritons}

After discussing the emission times of neutrons, protons, deuterons, and tritons from the Au + Au collisions at 0.4$A$ GeV, we now proceed the systematical analysis on correlation function for different particle pair combinations among neutrons, protons, deuterons, and tritons.  The correlation functions will be discussed  with specific gates on impact parameter, in-medium NNCS factor, total particle pair momentum as well as the particle velocity.
As mentioned in Sec. IIIA, our correlation functions are calculated by using the phase-space information from the freeze-out stage, which is used as the input for the Lednick$\acute{y}$ and Lyuboshitz code and then the effective source size is extracted by assuming a Gaussian-type emission source.

We first show in Fig.~\ref{fig5_ptot}  four types of identical light-particle correlation functions, namely $n-n, p-p, d-d,$ and $t-t$, for central collisions of $^{197}$Au + $^{197}$Au at E = 0.4$A$ GeV. The dependence of strength of the correlation functions on total particle pair momentum $\left(P_{tot}\right)$ will be discussed through the calculations with two gates on $P_{tot}$. In Fig.~\ref{fig5_ptot}, the curves with  open and filled circles are results with high $P_{tot}$ and low $P_{tot}$, respectively.
Fig.~\ref{fig5_ptot}(a) and (b) show the two-neutron and two-proton correlation functions for two different total momentum range of proton pairs (low: 0 - 0.4 GeV/c, high: 0.8 -1.2 GeV/c). Two-deuteron correlation function is presented in Fig.~\ref{fig5_ptot}(c) gated on the different total momentum of deuteron pairs (low: 0 - 0.8 GeV/c, high: 1.6 - 2.4 GeV/c). Shown in Fig.~\ref{fig5_ptot}(d) is two-triton correlation function gated on different total momentum of triton pairs (low: 0 - 1 GeV/c, high: 2 - 3 GeV/c).
From the figure, the shape of the correlation functions is consistent with those observed in experimental data from heavy-ion collisions \cite{Ghetti2000}. For neutron-neutron correlation function, it peaks at $q\approx 0$ MeV/c. The two-proton (b), two-deuteron (c) and two-triton (d) correlation functions are all suppressed at low $q$ because of Coulomb repulsion. The anti-symmetrization of the two-proton wave function may also suppresses low-$q$ pairs of protons, possibly enhancing this anti-correlation signal. With increasing relative momentum, for the two-proton correlation function, the strong final-state singlet-wave attraction gives rise to a maximum at $q \approx$ 20 MeV/c. However, the two-deuteron correlation function does not exhibit a peak since the anti-correlation between two-deuteron pairs induced by the repulsive singlet-wave nuclear potential and Coulomb potential. For the two-triton correlation function, it is also anti-correlated as shown in Fig.~\ref{fig5_ptot}(d) because  only the Coulomb potential is included  in the final-state interaction as in Ref. \cite{Boal1990}.
In Fig.~\ref{fig5_ptot}(a) and (b), it is clear observed that in the cut of higher $P_{tot}$, it leads to larger strength of the two-neutron and two-proton correlation functions. The trend implies that particles with higher momenta emitted earlier or equivalently from a compact source thus induces stronger correlation functions, consistent with the results shown in Fig.~\ref{fig4_emission-ekin}. 
The results are similar with a relatively simple approach in Refs.~\cite{Hagel2000} which has measured emission time for nucleons and light clusters in the coalescence model. The correlation between energy and emission time has been also clearly demonstrated in experimental data and model results for the momentum-gated nucleon pairs as demonstrated in Refs.~\cite{Gong1991,Colonna,Ma2006,Chen2003}. The momentum correlation function is very well complementary to above previous approach in terms of researching on the properties and the space-time evolution of reaction system.
 
However, the sensitivity to total pair momentum  becomes gradually weaker with increasing particle mass, eg. for deuterons and tritons.

Next we will see the effect of in-medium nucleon-nucleon cross section on momentum correlation function under different total pair momentum.  In Fig.~\ref{fig6_ptot-eta}, the curves with  filled and open circles are results with $\eta = 0.2$ and 0.5, respectively.
It  shows that the correlation functions of light particle pairs with high total momenta are more sensitive to the dependence of in-medium NNCS factor than that with the low total momenta at the same impact parameter. Since pre-equilibrium light particles with higher momenta are emitted earlier or have a smaller source size for smaller $\eta$. The $\eta$ dependence on  correlation functions with low total pair momenta shows opposite trend and this is consistent with the $\eta$ effect on particle emission times as shown in Fig.~\ref{fig4_emission-ekin}.

Fig.~\ref{fig7_radius} shows the dependence of radius extracted from light particle correlation functions on different in-medium NNCS factor and impact parameter for different total pair momentum gates, where the squares and triangles are results with low $P_{tot}$ and high $P_{tot}$ cut, respectively. The radii are extracted by a Gaussian source assumption, i.e.  $S\left(\textbf{r}^*\right)\approx\exp\left(-\textbf{r}^{*^{2}}/\left(4r_{0}^2\right)\right)$, where $r_{0}$ is the Gaussian source radius from the correlation functions. The results for light particle pairs without momentum-gated are shown by the curve with circles, which are of course in between the results of the high $P_{tot}$ and low $P_{tot}$ case. The extracted source radii from $p-p$ (b) and $n-n$ are similar but quantitatively different, which might be due to the effect of Coulomb distortions between proton-proton pair.  Source radii from $t-t$ (d) and $d-d$ (c) correlation functions are generally smaller than those extracted from $p-p$ (b) and $n-n$ (a).
 For $d-d$ and $t-t$ correlations as shown in Fig.~\ref{fig7_radius}, it is seen that the lower $\eta$, i.e. larger in-medium nucleon-nucleon cross section, leads to a slightly smaller radius, $i.e.$, stronger anti-correlation of deuteron or triton pairs than those obtained with the larger $\eta$, particularly for deuteron or triton pairs with high $P_{tot}$.
 For impact parameter dependence of the radii, we find that the radii general increase with impact parameter except in the low $P_{tot}$ case.

Finally, we also investigate the non-identical particle correlation functions, such as $p-d$, $p-t$, $d-t$ and $n-p$. The $p-d$ correlation function in Fig.~\ref{fig8_vpdt-ratio}(a) displays a single broad peak, due to both singlet-wave attraction and Coulomb repulsion. The shape is similar to the proton-proton correlation function while the peak is shown at about $q \approx$ 55 MeV/c. However, the $p-t$ and $d-t$ correlation functions in Fig.~\ref{fig8_vpdt-ratio}(a) are characterized by an anti-correlation due to final-state Coulomb repulsion. Due to the s-wave attraction it peaks at $q\approx 0$ MeV/c for neutron-proton correlation function. Except the non-identical particle correlation functions, the analysis of velocity-gated correlation functions of non-identical particles is a very powerful tool to probe detailed information about the particle emission time sequence \cite{Verde2006,Gelderloos1994,Gelderloos1995,Lednicky1996,Voloshin1997,Gelderloos1995prc,Ghetti2001}.
Fig.~\ref{fig8_vpdt-ratio}(b)  shows the ratios of  proton-deuteron, proton-triton, deuteron-triton, and neutron-proton correlation functions calculated with different velocity-gates. The ratio is defined by comparing two velocity-gated  correlation functions. The first function, $v+$ is constructed with pairs where the velocity of proton (deuteron) is faster than the deuteron (triton) or triton, respectively. The second function, $v-$ corresponds to the reverse situations. We obtain the emission sequences in nuclear collisions by a basic ideal as following: if one of the two particles is emitted earlier and owns lower velocity will, on average, travel shorter distances before the another particle is emitted. In our work, when the first emitted particle is slower than the second, the average distance will be reduced and the Coulomb suppression effect is thus enhanced, and vice versa. Therefore, in Fig.~\ref{fig8_vpdt-ratio}(b) the ratio of two different $p-d$ and $p-t$ correlation function which is lower than unity indicates that deuterons and tritons are, on the average, emitted earlier than protons in the low relative momentum region. However, the ratio of $n-p$ or $d-t$ correlation function which is just slightly higher or lower than unity, respectively. The phenomenon indicates that the difference of emission time between neutron and proton or deuteron and triton is not so significant. It is consistent with the emission time in  Fig.~\ref{fig3_emission-time}. On the contrary, those particles are emitted in the similar time scale in larger relative momentum region. The results are qualitatively consistent with other reaction systems, $^{36}$Ar + $^{27}$Al, $^{112}$Sn and $^{124}$Sn at 61$A$ MeV \cite{Ghetti_NPA2}.

\section{SUMMARY}

In summary, we present  results of particle-particle momentum correlation functions reconstructed by the Lednick$\acute{y}$ and Lyuboshitz analytical formalism using the phase-space points at the freeze-out stage for $^{197}$Au + $^{197}$Au collisions at different beam energies in a framework of the IQMD transport approach. As a necessary check for our model calculations,
we  performed a quantitative comparison  of proton-proton momentum correlation function with the FOPI data. Taking the same transverse momentum and rapidity phase space coverage corresponding to the experimental situation,  it is found that with increasing beam energy from 0.4$A$ GeV to 1.5$A$ GeV, the p-p correlation function becomes  stronger,
and they can well reproduce the FOPI experimental data of the proton-proton correlation functions. After this essential verification for  our model calculations, we can put forward for the following studies on emission time and momentum correlations of different light particles.

Emission time distributions of light particles  and their dependence on particles' c.m. kinetic energy are studied by taking two different in-medium NNCS and impact parameter sets.  We find that emission times are earlier for the particles with higher kinetic energies in central collisions. For semi-peripheral collisions, the average emission times of deuterons and tritons first increase with the kinetic energy and then drop.
At low kinetic energies, the larger in-medium nucleon-nucleon cross section makes the emission time longer, however, at higher kinetic energies, the effect becomes contrary. It indicates that the different emission origins, i.e. the lower kinetic energy particles are probably dominantly from statistical emission, and while the higher ones from pre-equilibrium dynamical process.

Momentum correlation functions with  total momentum  gated  for all different pairs of particles containing neutrons, protons, deuterons, and tritons have been investigated. The two-particle correlation functions, especially for neutron-neutron and proton-proton pairs with higher total momentum are stronger than the one with lower. The correlation function of light particle pairs and the emission source size gated on higher total momentum is sensitive to impact parameter and in-medium NN cross sections:  source size increases from central to semi-peripheral collisions, and source size becomes larger in the larger $\eta$ case, i.e. the smaller in-medium NN cross section.

Momentum correlations between non-identical light particles can provide important information about the emission sequence and the radius of their emitting sources. The results indicate that heavier clusters (deuterons or tritons) are emitted earlier than lighter ones at same momentum per nucleon as expected from the analysis of velocity-gated correlation functions of non-identical particles.

\section*{Acknowledgments}
This work is partially supported by the National Natural Science Foundation of China under Contract Nos. 11890714 and 11421505,  the Key Research Program of Frontier Sciences of the CAS under Grant No. QYZDJ-SSW-SLH002 and the Strategic Priority Research Program of the CAS under Grant No. XDPB09 and XDB16.

\end{CJK*}
\end{document}